\documentclass[
	aps, aps,pra,longbibliography, superscriptaddress, twocolumn,
	10pt
	floatfix, 
    nofootinbib,
	tightenlines
]{revtex4-1}
\usepackage[final]{graphicx}
\usepackage{times,bbm,amsmath,amssymb}
\usepackage{epsfig,color}
\usepackage{xcolor}
\usepackage{hyperref}
\hypersetup{
    colorlinks = true
}
\usepackage{cleveref}
\usepackage{microtype}
\usepackage{gensymb}
\usepackage{float,siunitx}
\usepackage[caption = false]{subfig}

\usepackage[greek,english]{babel}
\usepackage{thumbpdf,enumerate}
\usepackage{booktabs}
\usepackage{sidecap}
\usepackage[scaled=.8]{couriers}
\usepackage{multirow}
\usepackage{placeins}
\usepackage{relsize}
\usepackage{pst-grad,bm}
\usepackage{epigraph}
\usepackage{gensymb}
\usepackage{longtable}
\usepackage{ulem} 
\usepackage{acronym}
\usepackage{physics}
\usepackage{easyReview}

\usepackage[columnwise]{lineno}

\DeclareUnicodeCharacter{0301}{\'{e}}

\begin{document}

\title{Imaging Walk-Off--Driven Distortions in EPR Photon-Pair Correlations}
\author{Christian Howard}
\affiliation{Nexus for Quantum Technologies, University of Ottawa, K1N 5N6, Ottawa, ON, Canada}

\author{Roohollah Ghobadi}\email{farid.ghobadi80@gmail.com}
\affiliation{Nexus for Quantum Technologies, University of Ottawa, K1N 5N6, Ottawa, ON, Canada}

\author{Nazanin Dehghan}
\affiliation{Nexus for Quantum Technologies, University of Ottawa, K1N 5N6, Ottawa, ON, Canada}

\author{Alessio D'Errico}\email{aderrico@uottawa.ca}
\affiliation{Nexus for Quantum Technologies, University of Ottawa, K1N 5N6, Ottawa, ON, Canada}

\author{Ebrahim Karimi}
\affiliation{Nexus for Quantum Technologies, University of Ottawa, K1N 5N6, Ottawa, ON, Canada}
\affiliation{Institute for Quantum Studies, Chapman University, Orange, California 92866, USA}

\date{\today}

\begin{abstract}
Spontaneous parametric down-conversion is the primary source of position-correlated and momentum--anti-correlated photon pairs that form the canonical Einstein–Podolsky–Rosen (EPR) state. Their transverse spatial correlations are usually analyzed within the thin-crystal approximation, where the two-photon wavefunction is assumed to factorize into independent functions of the sum and difference coordinates. In practice, however, birefringence-induced transverse walk-off breaks this factorization and couples these degrees of freedom. Here, we show that this coupling persists even for nominally thin crystals once the free-space propagation of the joint spatial intensity is taken into account. This sum-difference coordinate coupling leads to a distinctive tapering of the transverse correlations near the crystal image plane--an effect that standard factorized models cannot capture. Numerical simulations and experimental data clearly confirm this novel behavior. Our findings provide a more complete description of photon-pair generation in birefringent nonlinear media and clarify fundamental limits on spatially resolved quantum imaging and spatial-mode quantum information processing with EPR states.
\end{abstract}

\maketitle

\section{Introduction}
Spontaneous parametric down-conversion (SPDC) is the most widely used source of entangled photon pairs and has enabled seminal demonstrations of quantum interference \cite{hong1987measurement}, entanglement generation \cite{shih1988new,kwiat1995new,weihs1998violation}, Einstein--Podolsky--Rosen (EPR) correlations \cite{howell2004realization}, and quantum steering \cite{saunders2010experimental}. By providing controllable, high-purity single photons and well-characterized entangled states, SPDC has transformed quantum entanglement from a foundational concept into a practical resource for quantum communication \cite{yuan2010entangled,scarfe2025}, imaging \cite{pittman1995optical,moreau2019imaging}, and photonic quantum information processing \cite{slussarenko2019photonic}. In this sense, SPDC serves as a versatile bridge between fundamental quantum optics and scalable photonic technologies \cite{o2009photonic}. \newline
The theoretical description of SPDC commonly relies on the thin-crystal approximation, in which the biphoton amplitude factorizes into contributions depending on the sum and difference of the transverse coordinates. This separation effectively decouples the pump-envelope function from the phase-matching response and underpins a wide range of analyses, including phase-retrieval and state-reconstruction protocols \cite{ortolano2023quantum,dehghan2024biphoton}, orbital angular momentum (OAM) conservation and engineering \cite{mair2001entanglement}, and entanglement migration between spatial degrees of freedom \cite{chan2007transverse}. When both the pump and phase-matching functions are further approximated as Gaussians, compact analytical expressions become available for quantifying spatial entanglement through the Schmidt number \cite{ekert1995entangled,parker2000entanglement,law2004analysis,miatto2012spatial} and experimentally accessible estimators such as the Fedorov ratio and EPR variances \cite{fedorov2006short,hong1987measurement,gomez2012quantifying}. These approximations are well justified in the thin-crystal, small-transverse-walk-off regime, where the phase-matching bandwidth is broad, the pump profile largely determines the joint spatial structure, and non-Gaussian distortions remain negligible.\newline
Free-space propagation, however, can substantially modify how spatial entanglement is distributed among the available degrees of freedom \cite{chan2007transverse,walborn2010spatial,tasca2008detection,tasca2009propagation,bhattacharjee2022propagation}. More recently, propagation has also been exploited as a diagnostic tool to reveal otherwise inaccessible phase information of the biphoton wavefunction and to enable its reconstruction \cite{dehghan2024biphoton}. Building on this perspective, we show here that free-space propagation can likewise expose birefringence-induced transverse walk-off effects through a coupling of the transverse sum and difference coordinates. Remarkably, this coupling can become observable and non-negligible under appropriate conditions, even for relatively thin nonlinear crystals ($L \approx 1$~mm). Specifically, we demonstrate that the resulting non-separability leads to an asymmetric and propagation-dependent broadening of the transverse spatial correlations. We further show that this effect is enhanced when the crystal is pumped with structured light, which amplifies the imprint of walk-off on the joint spatial correlations. We support these findings with heuristic arguments, numerical simulations, and direct experimental observations, and discuss their implications for accurately reconstructing and interpreting the spatial states of biphotons.

\section{Theory} 
\begin{figure*}[t]
  \centering
    \includegraphics[width=\textwidth]{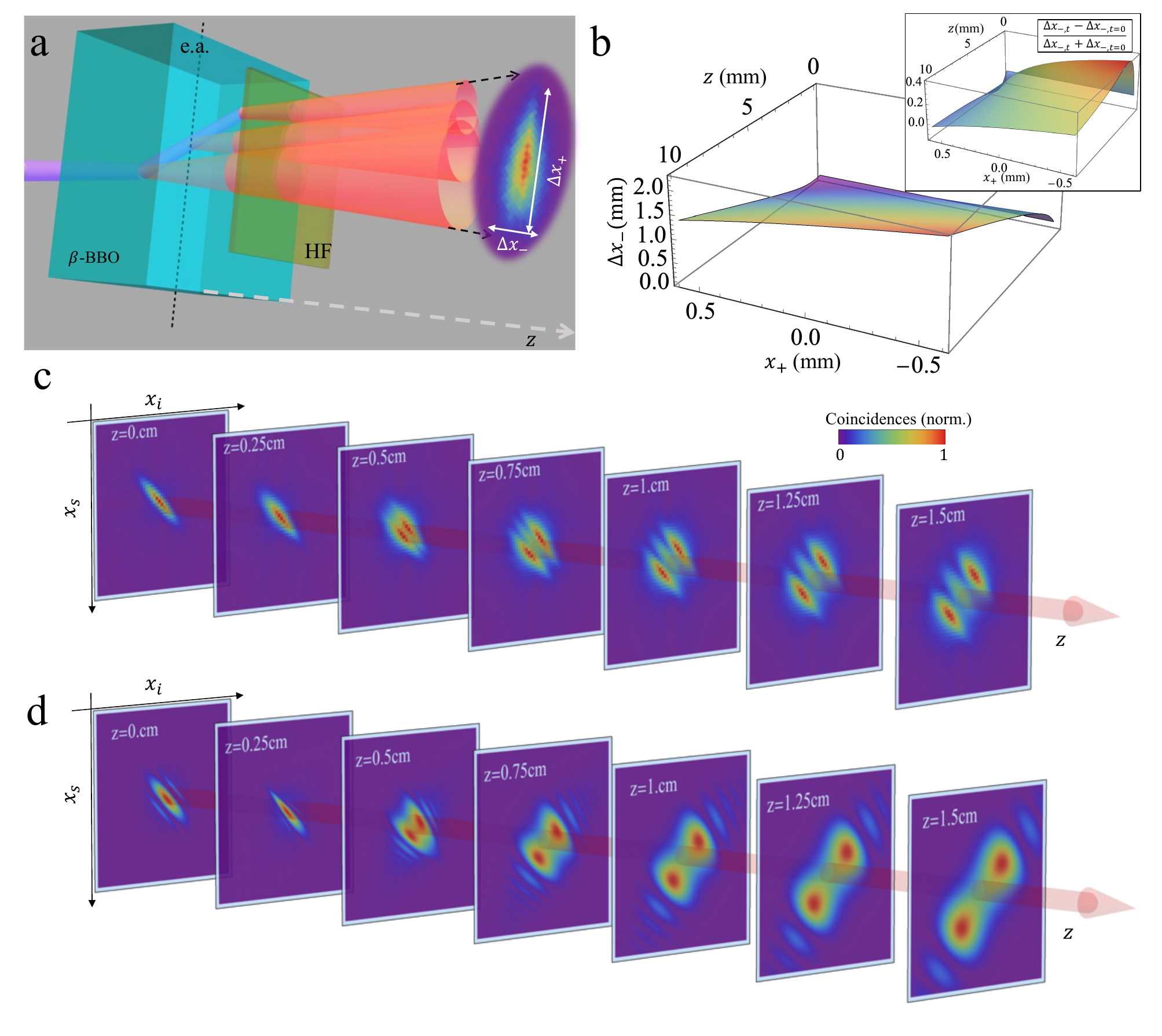}
    \caption{\textbf{Birefringence-induced transverse walk-off effects.} (a) Schematic illustration of birefringence-induced transverse walk-off in a nonlinear crystal. A pump beam (purple) with a polarization state along extraordinary axes of the crystal undergoes anomalous refraction, causing its centroid to drift laterally along the optic-axis direction as it propagates through the crystal. Consequently, photon pairs generated at different longitudinal positions emerge from laterally displaced locations, leading to a position-dependent width $\Delta x_-$ of the transverse anti-correlations. HF denotes a high-pass spatial filter. (b) Anti-correlation width $\Delta x_-$ as a function of propagation distance $z$ and sum coordinate $x_+$. The inset shows the normalized ratio $\Delta x_-(z,x_+)/\Delta x_-(z,x_+=0)$, highlighting that the dependence on $x_+$ is strongest near the crystal image plane. (c) Numerical simulations of the free-space evolution of the transverse spatial correlations in the presence of transverse walk-off, obtained from the propagation of the Wigner function expressed in the transformed sum--difference coordinate frame. (d) Corresponding simulations obtained using the angular spectrum propagation method, starting from an initial two-photon wavefunction given by the ansatz in Eq.~\eqref{eq:ansatz}.}
    \label{fig:theory}
\end{figure*}
We consider a uniaxial nonlinear crystal slab of thickness $L$, whose optic axis is cut at an angle $\theta$ with respect to the propagation direction $z$ of the pump beam. The two-photon wavefunction in the transverse-momentum representation, evaluated at the output face of the crystal ($z=L$), can be written as the product of a pump-envelope contribution and a phase-matching term. We assume a Gaussian pump profile of the form $e^{-w_p^2 \mathbf{p}_{+}^2/4}$, where $w_p$ denotes the pump waist at the crystal plane and $\mathbf{p}_{\pm}=(\mathbf{p}_s \pm \mathbf{p}_i)/\sqrt{2}$, with $\mathbf{p}_s$ and $\mathbf{p}_i$ the transverse momenta of the signal and idler photons, respectively.

The phase-matching function has the general form $e^{-i L \Delta p_z/2}\,\mathrm{sinc}(L \Delta p_z/2)$, where the longitudinal momentum mismatch $\Delta p_z$ can be expressed in terms of the transverse momentum variables. This leads to the biphoton wavefunction~\cite{walborn2010spatial}
\begin{align}\label{biphoton1}
	\Phi_{0}(\mathbf{p}_{+},\mathbf{p}_{-}) &=e^{-w_p^2 \mathbf{p}_{+}^2/4}\, e^{-i(\beta \mathbf{p}_{-}^2 + t p_{+,x} - l)} \nonumber \\
	&\quad \times \mathrm{sinc}(\beta \mathbf{p}_{-}^2 + t p_{+,x} - l),
\end{align}
where $\beta = L/(4 \eta_p k_p)$, with $\eta_p$ an effective refractive index (see Supplementary I and Ref.~\cite{walborn2010spatial}). The parameters $t$ and $l$ represent the transverse and longitudinal walk-off terms, respectively. The longitudinal walk-off arises from the different group velocities of the pump and the down-converted photons, while the transverse walk-off---which is central to this work---originates from the anomalous refraction of the pump beam inside the birefringent crystal, here assumed to be extraordinarily polarized.

A key point is that finite transverse walk-off introduces an explicit dependence of the phase-matching function on the sum coordinate $p_{+,x}$. In the absence of this contribution, the phase matching would depend solely on $\mathbf{p}_-$, and the biphoton wavefunction would factorize in the sum--difference representation as $\Phi_0(\mathbf{p}_{+},\mathbf{p}_{-}) = A_p(\mathbf{p}_{+})\,\psi(\mathbf{p}_{-})$. This separability is a common assumption and is often taken to be valid in the thin-crystal limit~\cite{walborn2010spatial}. Only a few works have explicitly addressed the effects of transverse walk-off. In Ref.~\cite{moura2024einstein}, transverse walk-off was incorporated by approximating the phase-matching function as $\mathrm{sinc}(\beta \mathbf{p}_{-}^2 + t p_{+,x}) \approx \mathrm{sinc}(\beta \mathbf{p}_{-}^2)\,\mathrm{sinc}(t p_{+,x})$. For relatively thick crystals ($L=5~\mathrm{mm}$), this approach was shown to accurately describe far-field two-photon correlations, while failing in the near field. Importantly, however, this approximation restores separability between the sum and difference coordinates. In contrast, we show that the intrinsic coupling between $\mathbf{p}_+$ and $\mathbf{p}_-$ can play a significant role even for relatively thin crystals ($L \sim 1~\mathrm{mm}$), where it manifests as pronounced asymmetries in the spatial correlations.
 
We begin with an intuitive picture that illustrates how coupling in the phase-matching term arises. As sketched in Fig.~\ref{fig:theory}-(a), the extraordinary pump undergoes transverse walk-off inside the crystal, drifting sideways along the optic-axis direction. As a result, photon pairs generated at different longitudinal positions exit the crystal from laterally displaced launch points. Upon free-space propagation to the detection plane, these depth-dependent displacements overlap to form an asymmetric envelope. On the walk-off side, photon pairs generated later in the crystal accumulate into a sharp, tapered edge, whereas on the opposite side, pairs created earlier or near the middle of the crystal spread out more broadly, giving rise to the observed wedge-shaped correlations.

This behavior can be understood through a simple heuristic argument. The width of the transverse anti-correlation, $\Delta x_-(z)$, measured at a propagation distance $z$ from the crystal, depends on both the initial position width $\Delta x_{-}(0)$ and the momentum width $\Delta p_{-}(0)$ evaluated at the crystal plane. For free-space propagation, this dependence follows directly from the ABCD law of geometrical optics, which, expressed in terms of the sum and difference transverse variables $\mathbf{x}_{\pm} = (\mathbf{x}_s \pm \mathbf{x}_i)/\sqrt{2}$, reads~\cite{fowles1989introduction}
\begin{equation}
	\begin{pmatrix} 
	\mathbf{x}_{\pm}(z) \\
	\mathbf{p}_{\pm}(z)
	\end{pmatrix}
=
	\begin{pmatrix}
	1 & z/k \\
	0 & 1
	\end{pmatrix}
	\begin{pmatrix}
	\mathbf{x}_{\pm}(0) \\
	\mathbf{p}_{\pm}(0)
	\end{pmatrix},
\label{Hpm}
\end{equation}
where $k = k_s = k_i$ denotes the signal and idler wavenumbers. From this relation, the propagation of the anti-correlation width can be written as
\begin{equation}
	\Delta x_-(z) = \sqrt{[\Delta x_-(0)]^2 + (z/k)^2 [\Delta p_-(0)]^2 }.
	\label{eq:spread1}
\end{equation}

To estimate $\Delta p_-(0)$, we examine the phase-matching kernel $\mathrm{sinc}(\beta p_-^{2} + t p_{+,x} - l)$, which explicitly couples the even $p_-^{2}$ term to a linear bias in the sum momentum $p_{+,x}$. Approximating the momentum width by the location of the first zero of the sinc function yields $\Delta p_-(0) = 2\sqrt{(\pi + l - t p_{+,x})/\beta}$. Substituting this into Eq.~\eqref{eq:spread1} gives
\begin{equation}
	\Delta x_-(z) =\sqrt{[\Delta x_-(0)]^2 + \frac{4 z^2}{k^2 \beta} (\pi + l - t p_{+,x})}.
	\label{eq:spread2}
\end{equation}

Using Eq.~\eqref{Hpm}, the sum momentum can be related to the sum coordinate through $p_{+,x} = k[x_+(z) - x_+(0)]/z$. Estimating $x_+(0) \sim w_p/2$ leads to the final expression
\begin{equation}
	\Delta x_-(z,x_+) =w_{\phi}\sqrt{1 + \frac{z^2}{z_{\mathrm{eff}}^2}
	\left(\pi + l - t k \frac{x_+ - w_p}{z}\right)},
\end{equation}
where $w_{\phi} \equiv \Delta x_-(0)$ and $z_{\mathrm{eff}} = k \sqrt{\beta}\, w_{\phi}/2$.

In the Gaussian-approximation limit, where walk-off is neglected ($t = 0$, $l \approx 0$) and $\sqrt{\beta} \sim w_{\phi}$, this expression reduces to the familiar divergence law of Gaussian beams, $w_{\phi}(z) = w_{\phi} \sqrt{1 + (z/z_{\mathrm{eff}})^2}$. Finite transverse walk-off introduces an additional correction to the squared anti-correlation width that depends explicitly on the sum coordinate $x_+$ and grows linearly with propagation distance $z$ to first order, as illustrated in Fig.~\ref{fig:theory}-(b). This correction is responsible for the characteristic wedge-shaped spatial correlations observed experimentally. At large propagation distances, the usual $(z/z_{\mathrm{eff}})^2$ term dominates, rendering the walk-off-induced asymmetry less pronounced. This is evident in the inset of Fig.~\ref{fig:theory}-(b), where we plot the normalized difference $(\Delta x_{-,t} - \Delta x_{-,t=0})/(\Delta x_{-,t} + \Delta x_{-,t=0})$, showing that the effect of finite transverse walk-off is strongest near the crystal image plane. We emphasize that this treatment is qualitative in nature, as it does not fully account for the detailed propagation effects arising from the sinc-shaped phase-matching function.

\begin{figure*}
    \centering
        \includegraphics[width=\textwidth]{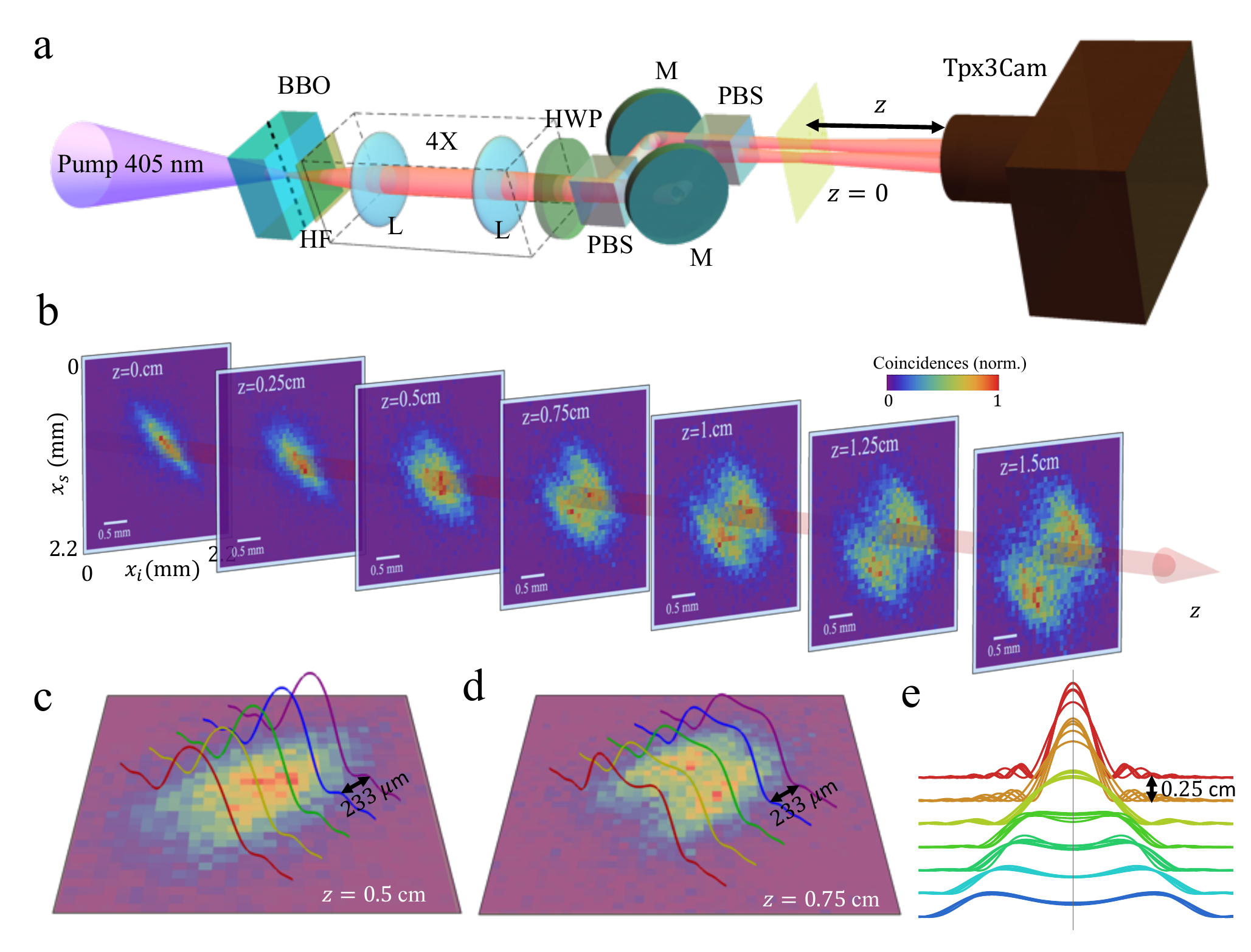}
        \caption{\textbf{Experimental results.} (a) Schematic of the experimental setup. BBO: 1-mm-thick type-I $\beta$-barium borate crystal; HF: high-pass filter; L: lens; HWP: half-wave plate; PBS: polarizing beamsplitter; M: mirror. (b) Experimentally reconstructed transverse spatial correlations along the $x$ coordinate. (c) Corresponding simulation results obtained using the ansatz in Eq.~\eqref{eq:ansatz}. (d,e) Detailed analysis of transverse walk-off effects, highlighting the dependence of the correlation width on the sum coordinate $x_+$. The one-dimensional cuts show best fits of the coincidence distributions for fixed values of $x_i - x_s$, sampled in steps of $233~\mu$m, illustrating how the spatial correlations evolve along this direction due to transverse walk-off. (f) Same one-dimensional analysis as in (d,e), shown for all measured propagation planes. When stacked together, these reconstructed profiles map out the free-space propagation of the phase-matching function.}
         \label{fig:exp}
\end{figure*}

To obtain more quantitative insight, we perform numerical simulations using two complementary approaches. In both cases, to keep the problem computationally controllable, we restrict the analysis to a single transverse dimension and neglect the dependence on the $y$ coordinate.

The first approach is based on calculating the Wigner function of the biphoton state, defined as~\cite{leonhardt1997measuring}
\begin{eqnarray}\label{wigner} \nonumber
	W(\mathbf{x}_{+},\mathbf{x}_{-};\mathbf{p}_{+},\mathbf{p}_{-}) &=& \frac{1}{(2\pi)^4}
	\iint d^2{q}_{+}\, d^2{q}_{-}\,e^{i(\mathbf{x}_{+}\mathbf{q}_{+}+\mathbf{x}_{-}\mathbf{q}_{-})} \quad \\ \nonumber
	&\times&
	\Phi_0\!\left(\mathbf{p}_{+}+\frac{\mathbf{q}_{+}}{2},\mathbf{p}_{-}+\frac{\mathbf{q}_{-}}{2}\right) \\ 
	&\times&
	\Phi_0^{*}\!\left(\mathbf{p}_{+}-\frac{\mathbf{q}_{+}}{2},\mathbf{p}_{-}-\frac{\mathbf{q}_{-}}{2}\right).
\end{eqnarray}
Under the linear transformation in Eq.~\eqref{Hpm}, the free-space evolution of the Wigner function takes a particularly simple form \cite{leonhardt1997measuring,alonso2011wigner}
\begin{equation}\label{Wprop}
	W(\mathbf{x}_{\pm}(z),\mathbf{p}_{\pm}(z)) =
	W\!\left(\mathbf{x}_{\pm}-\frac{z}{k}\mathbf{p}_{\pm},\mathbf{p}_{\pm}\right).
\end{equation}

In our simulations, Eq.~\eqref{Wprop} is used to propagate the Wigner function through free space. Specifically, we first calculate the initial Wigner function corresponding to the biphoton state in Eq.~\eqref{biphoton1}, using the parameters $w_p = 50$, $\beta = 1$, $t = 4.0$, and $l = 3.0$. The propagation is implemented by transforming the phase-space axes and interpolating the Wigner function onto the new coordinate grid. By integrating $W(p_i(z),p_s(z),x_i(z),x_s(z))$ over the momentum variables $(p_i(z),p_s(z))$, we obtain the position-space marginal probability distribution, which is proportional to the experimentally measured spatial correlations at distance $z$.

The resulting transverse spatial correlations for several propagation planes are shown in Fig.~\ref{fig:theory}-(c). The simulations reveal the characteristic emergence of a right-diagonal lobe, whose broader end is aligned along the line $x_i = x_s$ and which becomes increasingly pronounced with propagation distance.

As a second approach, we adopt an ansatz for the near-field biphoton wavefunction in which the usual double-Gaussian approximation is supplemented by a phase-curvature term that explicitly couples the sum and difference coordinates. Guided by the physical picture in Fig.~\ref{fig:theory}-(a), we assume the following form:
\begin{align}
	\psi(x_+,x_-) =e^{i\frac{x_-^2 (x_+ - x_0)}{\varrho^3}}\,\mathrm{sinc}\!\left(\frac{x_-^2}{w_{\phi}^2}-\xi\right)e^{-x_+^2/w_p^2},
\label{eq:ansatz}
\end{align}
where the phase factor $e^{i x_-^2 (x_+ - x_0)/\varrho^3}$ captures the effects of transverse walk-off through an $x_+$-dependent curvature. The parameters $(\varrho, x_0, \xi)$ are treated as free variables and adjusted to obtain agreement with the experimental observations.

At the output face of the crystal, although the spatial correlations remain narrow, phase-curvature effects can already be present due to the propagation of the phase-matching function within the crystal. Importantly, the accumulated curvature depends on $x_+$: larger values of $x_+$ correspond to shorter effective propagation distances for the associated phase-matching cones. Drawing an analogy with the curvature of a TEM$_{00}$ Gaussian mode, for which $R(z) = z + z_R^2/z \approx z_R^2/z$ in the small-$z$ limit, we make the substitution $z \rightarrow (t/L)(x_+ - x_0) \ll 1$. This yields the estimate $\varrho^3 \sim k w_{\phi}^4 L / (2t)$, where $w_{\phi}$ is interpreted as the effective near-field width of the phase-matching function and $z_R \rightarrow k w_{\phi}^2/2$.

The parameters $\varrho$, $x_0$, and $\xi$ were optimized to match the experimental data, yielding $(\varrho, x_0, \xi) \approx (350~\mu\mathrm{m}, 5 w_p, -2)$. The values of $w_{\phi}$ and $w_p$ were independently extracted from the measured correlation widths along the $x_{\pm}$ directions. The corresponding simulation results are shown in Fig.~\ref{fig:theory}-(d), where the formation of the characteristic wedge-shaped correlations at intermediate propagation distances is again clearly observed.

Finally, we note that these simulations neglect coupling to the orthogonal transverse coordinate $y$ during propagation. As a result, while they capture the essential physics of transverse walk-off and its impact on spatial correlations, they cannot reproduce all quantitative features of the experimental data.

\section{Experiment}
\begin{figure}[t]
  \centering
    \includegraphics[width=0.5\textwidth]{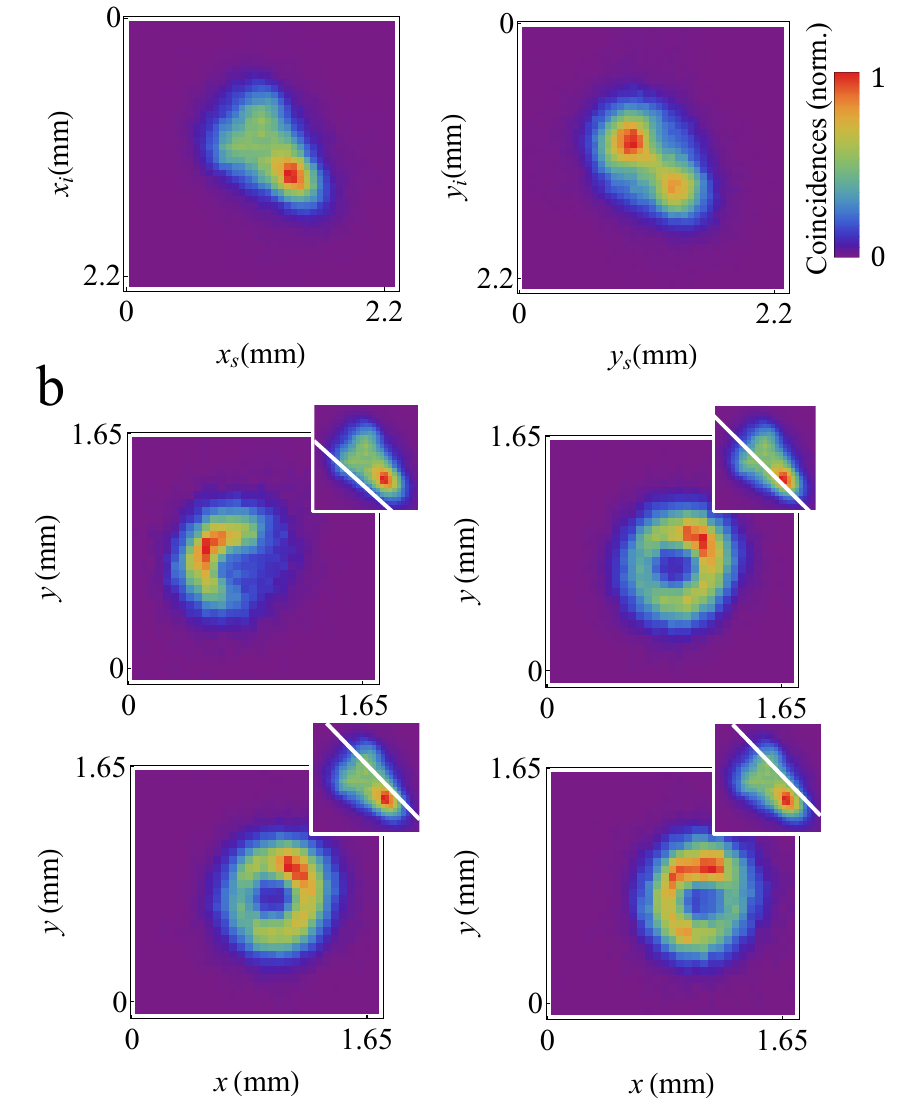}
    \caption{\textbf{Transverse walk-off effects with OAM pumping.} (a) Measured transverse spatial correlations at a propagation distance of $z=20$ mm when the crystal is pumped with a beam carrying orbital angular momentum $\ell = 6$. Transverse walk-off effects are particularly evident in the $x$-correlations, where they introduce pronounced asymmetries. (b) Coincidence images obtained by post-selecting on spatially correlated photon pairs satisfying $x_i = x_s + c$. In the absence of transverse walk-off, this procedure would directly reconstruct the pump intensity profile. When a walk-off is present, however, the reconstructed intensity becomes dependent on the post-selection condition, as highlighted in the insets. The rectangular bars indicate the different values of $c$ used for post-selection.}
    \label{fig:figoam}
\end{figure}

We experimentally verified the predicted effects of transverse walk-off by measuring the evolution of spatial correlations in SPDC generated from a $1$-mm-thick type-I BBO crystal pumped by a pulsed $405$-nm laser. The pump beam was focused onto the crystal, with a measured waist of approximately $0.2$ mm at the crystal plane, which satisfies the thin crystal approximation, i.e. $L/z_R\simeq 3\times10^{-3}$. The experimental setup is schematically shown in Fig.~\ref{fig:exp}-(a). After the crystal, an imaging system with $4\times$ magnification was used to measure near-field spatial correlations using an event-based camera (Tpx3Cam). Signal and idler photons were spatially separated, with $50\%$ probability, so as to impinge on two distinct regions of the camera sensor. This separation was achieved using a half-wave plate rotated by $22.5^\circ$ followed by two polarizing beam splitters, arranged such that both photons propagate over equal optical path lengths (see Ref.~\cite{dehghan2024biphoton} for further details). Spatial correlations were recorded at different propagation distances $z$, referenced to the image plane of the crystal ($z=0$). Measurements were performed for propagation distances ranging from $z=0$ to $15$ mm in steps of $2.5$ mm. The experimentally reconstructed spatial correlations clearly reveal the characteristic wedge-shaped structure predicted by our model (see Fig.~\ref{fig:exp}-(b)), which is particularly pronounced at intermediate propagation distances. Overall, the experimental observations are in good agreement with the numerical simulations.
A more detailed analysis of selected cases is presented in Fig.~\ref{fig:exp}-(d) and (e), corresponding to propagation distances of $z=5$ mm and $z=7.5$ mm, respectively. The colored curves represent fits to one-dimensional sections of the coincidence distributions taken along lines defined by $x_i = -x_s + c$, where $c$ is a variable offset parameter. The fits were performed using a $\mathrm{sinc}^2(A x^2 - B)$ functional form, with $A$ and $B$ as fitting parameters. Although this expression does not provide an exact analytical description of the phase-matching function at intermediate propagation planes, it serves as an effective phenomenological model (see Supplementary Fig.~2), as it can reproduce both Gaussian-like and double-lobed profiles.
In Fig.~\ref{fig:exp}-(d), the progressive broadening of the fitted sections with increasing $c$--corresponding to a transition from the purple to the red curves--directly reflects the influence of transverse walk-off. In Fig.~\ref{fig:exp}-(e), one can additionally observe the emergence of a pronounced double-lobed structure for larger values of $c$. This evolution closely resembles the free-space propagation of the phase-matching function in the absence of transverse walk-off ($t=0$), an analogy that becomes more evident when comparing measurements taken at different propagation planes, as shown in Fig.~\ref{fig:exp}-(f).
Transverse walk-off effects become even more apparent when the crystal is illuminated with structured pump beams. As a representative example, we consider a pump beam carrying an azimuthal phase factor $e^{i\ell\phi}$, corresponding to an orbital angular momentum (OAM) of $\ell\hbar$ per photon. Figure~\ref{fig:figoam}-(a) shows the measured spatial correlations at a propagation distance of $20$ mm from the crystal image plane, where a distinctly triangular structure emerges in the $x$-correlations.
In the absence of transverse walk-off, the pump spatial profile can be reconstructed by post-selecting on correlated photon pairs \cite{dehghan2024biphoton}. This is possible because the separability of the biphoton wavefunction in the sum and difference coordinates allows the pump and phase-matching contributions to be independently extracted from coincidence measurements. When transverse walk-off is non-negligible, however, information about the pump profile is not lost; instead, it becomes intrinsically coupled to the phase-matching function. Consequently, post-selection on spatially correlated events defined by $x_i = x_s + c$ yields results that depend explicitly on the chosen value of $c$, as illustrated in Fig.~\ref{fig:figoam}-(b). In all cases, the characteristic doughnut-shaped intensity profile associated with OAM beams remains visible, but it acquires a pronounced asymmetry due to its convolution with the phase-matching response.

\section{Conclusions}
In conclusion, we have shown that transverse walk-off induces an intrinsic coupling between the pump and phase-matching contributions, leading to asymmetric and propagation-dependent modulations of the joint position-space coincidence distribution. The resulting wedge-shaped spatial correlations constitute a previously unexplored effect that must be taken into account in quantum imaging and state-reconstruction experiments that rely on propagation-dependent measurements. Although transverse walk-off can degrade the direct interpretation of spatial correlations, our results demonstrate that a careful analysis based on coincidence post-selection can still be used to retrieve meaningful information about the constituent functions of the two-photon state. An interesting direction for future work is the investigation of these effects in the high-gain parametric down-conversion regime, where multimode dynamics may further enhance or modify the observed behavior.

More broadly, the approach presented here can be generalized to other scenarios in which the separability of the biphoton wavefunction in sum and difference coordinates breaks down during propagation \cite{di2009near,baghdasaryan2022maximizing,moura2024einstein}. Relevant examples include diffraction through out-of-focus objects, propagation through turbulent or scattering media \cite{jha2010effects,cameron2024adaptive}, and tight-focusing regimes where the paraxial approximation is no longer valid \cite{vega2022fundamental}. In these contexts, propagation-induced coupling between spatial degrees of freedom may provide both challenges and new opportunities for controlling and exploiting high-dimensional quantum correlations.

\subsection*{Acknowledgment}
The authors acknowledge discussion with Yingwen Zhang. 

\subsection*{Funding}
This work was supported by the Canada Research Chairs (CRC), National Research Council Canada Quantum Sensing Program, and Quantum Enhanced Sensing, Imaging (QuEnSI) Alliance Consortia Quantum grant. 
\subsection*{Author contributions} R.G., A.D. and C.H. developed the theory. C.H. and A.D. performed the numerical simulations. N.D. and A.D. prepared the experimental setup. A.D., with contributions from C.H. and R.G., performed the experiment and analyzed the data. N.D. and A.D. collected the data for the OAM pump. E.K. supervised the project. R.G., A.D. and C.H. prepared the first version of the manuscript. 
\subsection*{Competing interest}
The authors declare that they have no competing interests.
\subsection*{Data and materials availability}
All data are available upon reasonable request to the corresponding authors.
\bibliographystyle{apsrev4-1}
\bibliography{bibliography.bib}
\clearpage
\onecolumngrid
\renewcommand{\figurename}{\textbf{Figure}}
\setcounter{figure}{0} \renewcommand{\thefigure}{\textbf{S{\arabic{figure}}}}
\setcounter{table}{0} \renewcommand{\thetable}{S\arabic{table}}
\setcounter{section}{0} \renewcommand{\thesection}{S\arabic{section}}
\setcounter{equation}{0} \renewcommand{\theequation}{S\arabic{equation}}
\onecolumngrid

\begin{center}
{\Large Supplementary Material for: \\Imaging deformations of two-photon spatial correlations induced by transverse walk-off}
\end{center}
\vspace{1 EM}
\section{Dependence of biphoton wavefunction parameters on refractive indices and crystal orientation}
The parameters $t$, $l$ and $\eta_p$ in Eq. \ref{biphoton1} can be expressed in terms of refractive indices and the crystal orientation $\theta$ as:
\begin{align}
t&=\frac{L(n_o^2-n_e^2)\sin{\theta}\cos{\theta}}{2n^2(\theta)},\cr l&=\frac{Lk_p(\eta_p-\bar{n})}{2},\cr
\eta_p&=\frac{n_o n_e}{n(\theta)},
\end{align}
with $n^2(\theta)=n_o^2\sin^2{\theta}+n_e^2\cos^2{\theta}$,$\bar{n}=(n_o+n_e)/2$, and $n_o$ and $n_e$ are the ordinary and extraordinary refractive indices, respectively, at pump wavelength. In what follows, we focus on frequency degenerate SPDC, wherein the signal and idler have identical frequencies; the results extend straightforwardly to the non-degenerate regime.  
\section{Simulation Details.}
 The numerical computation of the Wigner function for the biphoton state is performed in Python using the NumPy and Matplotlib libraries. In order to compute the Wigner function we take the Fourier Transform of the argument $\Phi(\mathbf{p} - \frac{\mathbf{p}_0}{2}) \Phi^* (\mathbf{p} + \frac{\mathbf{p}_0}{2})$ with respect to $\mathbf{p_0}$. Therefore, to model the propagation of the biphoton Wigner function, we numerically compute the Wigner function using iterated Fast Fourier Transforms. We now present an outline of the program's construction. Pick a real number $P>0$ and consider the set S = $(-P,P)\times (-P,P) \subset \mathbb{R} \times \mathbb{R}$. The real number $P$, should be large enough so that the support of $\Phi(\mathbf{p} - \frac{\mathbf{p}_0}{2}) \Phi^* (\mathbf{p} + \frac{\mathbf{p}_0}{2})$ is contained in S. We then construct a uniform partition $\mathcal{P}$ of the set S into $N\times N$ cells. We then sample the midpoint of each of the cells in $\mathcal{P}$ to get the set of sampled momentum space coordinates $s = \{(p_{s_n}, p_{i_m})=(-P + \frac{P(1+2n)}{N}, -P + \frac{P(1+2m)}{N} ): n,m = 0,..., N-1\}$. Then for each fixed $\mathbf{p}_{n,m} = (p_{s_n}, p_{i_m}) \in s$, we compute the Fast Fourier Transform to get $W(x_s,x_i,p_{s_n},p_{i_m}) = FFT \{\Phi(\mathbf{p}_{n,m}- \frac{\mathbf{p}_0}{2}) \Phi^* (\mathbf{p}_{n,m} + \frac{\mathbf{p}_0}{2})\}$. The numerical Wigner function is then constructed by taking the set of all such $W(x_s,x_i,p_{s_n},p_{i_m})$. Hence the numerical Wigner function is the set $W(x_s,x_i,p_s,p_i) = \{W(x_s,x_i,p_{s_n},p_{i_m}) : n,m = 0, ... , N-1\}$. We realize the propagation of our Wigner function through a passive transformation of the position and momentum axes. Finally, we obtain our propagated position marginal distribution by numerically integrating over the momentum space variables $(p_s,p_i)$.\\

In our simulations, the momentum space $\{\mathbf{p}_0 : \mathbf{p}_0 = (p_{s_0},p_{i_0})\}$ is discretized into an $M\times M$ grid centered about the origin where $M=124$. The position space, $\{\mathbf{x}: \mathbf{x} = (x_s,x_i)\}$, which corresponds to the Fourier plane, will also be discretized into an $M \times M$ grid by the Fast Fourier Transform. The momentum space, $\{ \mathbf{p}:  \mathbf{p} =(p_s,p_i)\}$ is discretized into an $N\times N$ grid where $N=1.5 M = 186$. The propagation parameter, $\mu$, is allowed to take on the following set of values $\{0.0,0.25,0.5,0.75,1.0,1.25,1.50\}$. We observe that for $\mu=0$, which corresponds to the position marginal at the exit face of the crystal, we see very tight correlations along the main diagonal. In the case for zero transverse walk-off ($t=0$) we see that as we propagate the transverse plane, the position marginal transitions from being dominant along the correlation diagonal to becoming dominant along the anti-correlation diagonal. In the simulations with non-zero transverse walk-off  ($t=4.0$) and longitudinal walk-off ($l=3.0$), we observe the characteristic formation of a diagonal pointing lobe. As the propagation distance increases, we observe fictitious fringes that are simple artifacts of a lack of resolution in the position and momentum spaces. On the other hand, in the simulations with walk-off, we see that upon propagation, our position marginal distribution obtains a lobe shape whose wide end is directed along the correlation diagonal $x_i=x_s$. This means initially, we are nearly certain to detect the two photons having the same position, which makes sense since photons are generated from the same pump photon at a specific location within the BBO crystal. Then, upon propagation, we observe the transverse walk-off as it becomes more likely to detect the two photons farther apart from one another.

\section{Supplementary Figures}
Figure S1 compares the experimental data with the corresponding best fits that are shown in Fig. \ref{fig:exp} c-e for all propagation planes and transverse sections considered.

\begin{figure*}
    \centering
        \includegraphics[width=0.8\textwidth]{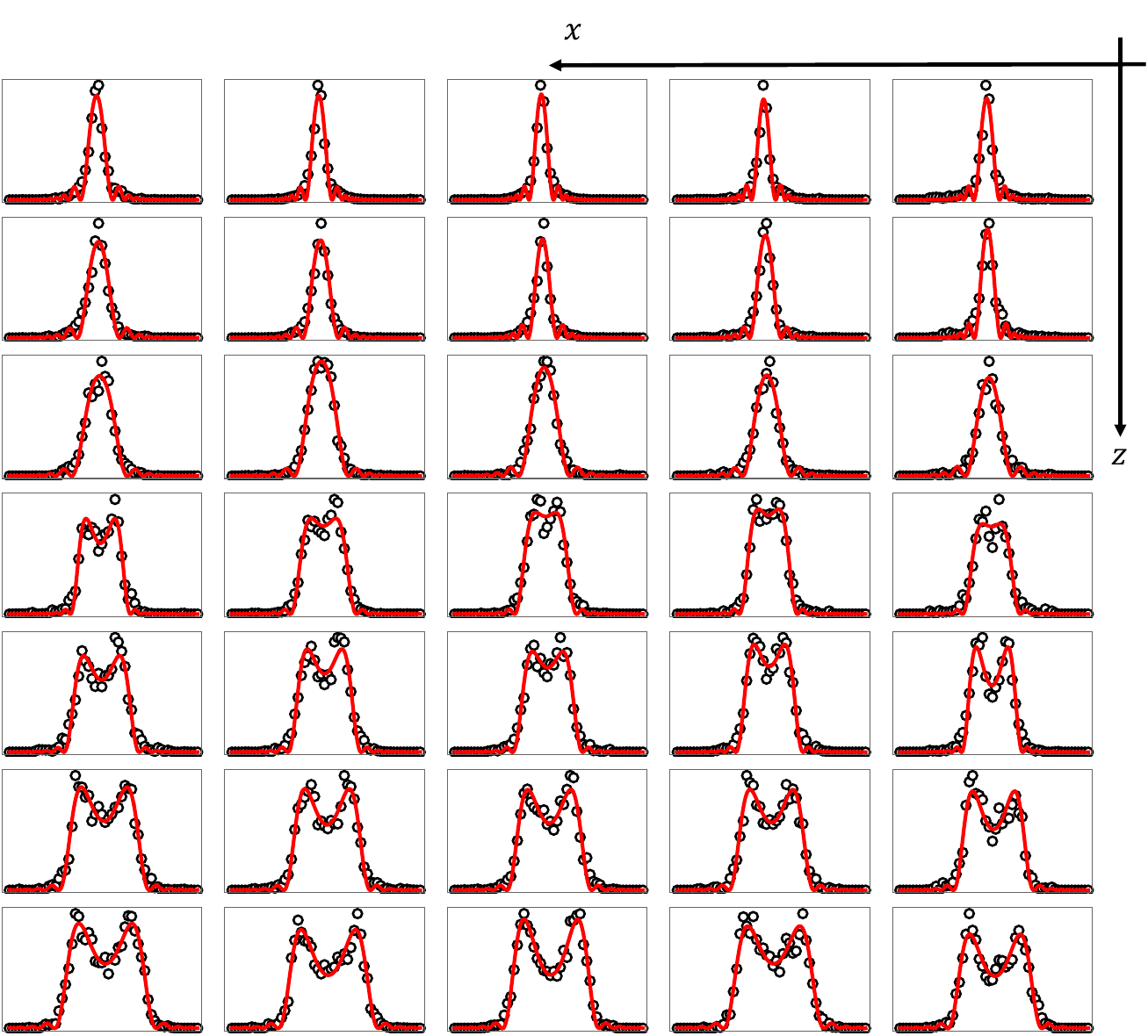}
        \caption{\textbf{Fits of sections of measured correlation patterns.} Full comparison between experimental data (black circles) and best fits (red plots) that was shown in Fig. \ref{fig:exp}-(c)-(e) for all propagation planes and sections considered.}
\end{figure*}

\end{document}